\PassOptionsToPackage{table,svgnames}{xcolor}
\documentclass{vgtc}                          

\usepackage{microtype}
\usepackage{pdfpages}

\graphicspath{{figures/}{pictures/}{images/}{./}} 

\usepackage{times}                     

\onlineid{0}

\vgtccategory{Research}

\vgtcinsertpkg

\title{Dissecting Atomic Facts: Visual Analytics for Improving Fact Annotations in Language Model Evaluation}

\author{Manuel Schmidt \thanks{e-mail: manuel.schmidt@uni.kn}\\ %
        \scriptsize University of Konstanz %
\and Daniel A. Keim \thanks{e-mail: daniel.keim@uni.kn}\\ %
    \scriptsize University of Konstanz
\and Frederik L. Dennig \thanks{e-mail: frederik.dennig@uni.kn}\\ %
     \scriptsize University of Konstanz %
}
\teaser{
  \centering
  \includegraphics[width=\linewidth]{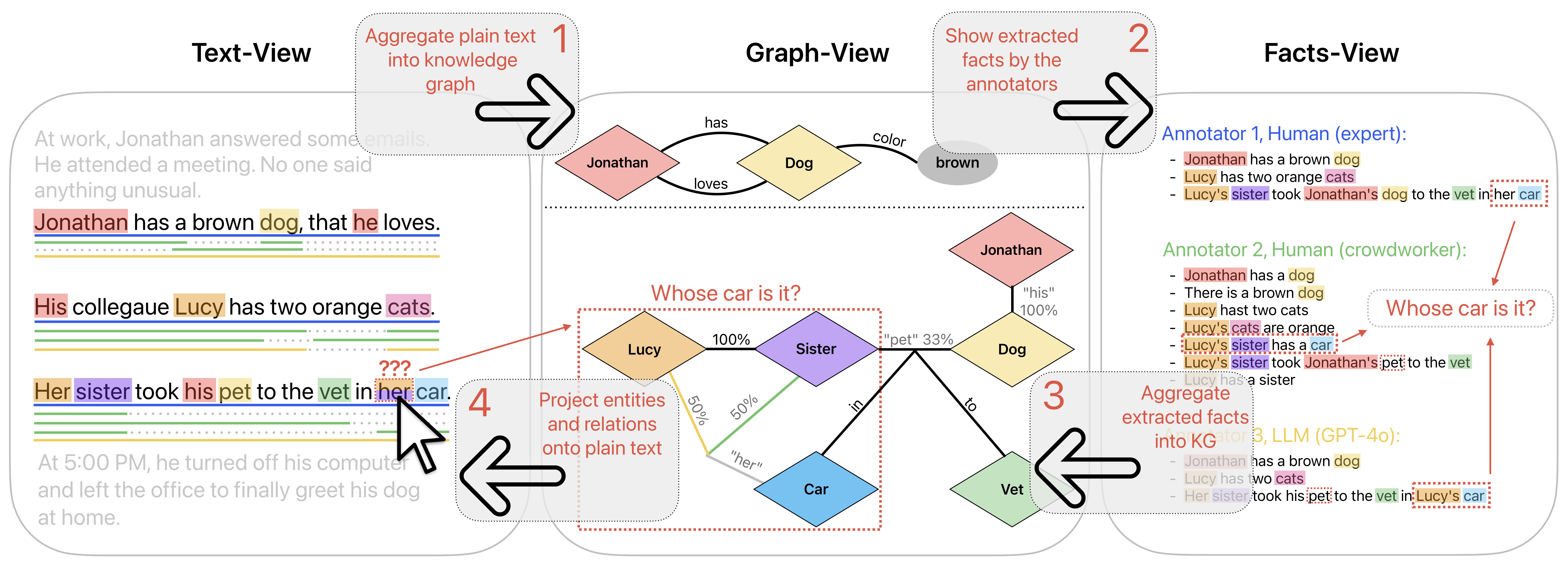}
  \caption{Whose car is it? Our conceptual design illustrates a visual interface revealing inconsistencies in factual annotations across human annotators and language models. Entity highlighting, atomic facts, and corresponding knowledge graphs are shown side by side. Ambiguous references, e.g., the unresolved ownership of ''her car'', are selected for inspection. Small-multiple graphs enable interactive exploration of the local context, providing insight into alternative interpretations and annotation uncertainty.}
  \label{fig:teaser}
}

\abstract{
    Factuality evaluation of large language model (LLM) outputs requires decomposing text into discrete “atomic” facts. However, existing definitions of atomicity are underspecified, with empirical results showing high disagreement among annotators, both human and model-based, due to unresolved ambiguity in fact decomposition.
    We present a visual analytics concept to expose and analyze annotation inconsistencies in fact extraction. By visualizing semantic alignment, granularity and referential dependencies, our approach aims to enable systematic inspection of extracted facts and facilitate convergence through guided revision loops, establishing a more stable foundation for factuality evaluation benchmarks and improving LLM evaluation.

} 


\begin{document}


\firstsection{Introduction}

\maketitle

Recent advances in large language models (LLMs) led to increased reliance on evaluating their \emph{factuality}, which describes the LLMs' ability to generate outputs that are accurate, verifiable, and coherent.
Validating an LLMs factuality is crucial for critical use cases in which reliable outputs are mandatory, such as in public services, healthcare or law.
However, evaluation pipelines such as \textit{FActScore} \cite{min-etal-2023-factscore}, \textit{TruthfulQA} \cite{lin-etal-2022-truthfulqa}, or \textit{FEVER} \cite{thorne-etal-2018-fever} rely heavily on extracted atomic facts or claims: minimal, discrete units of meaning against which outputs can be evaluated. 
This raises a foundational issue: What constitutes an atomic fact?\\
Despite being central to factuality evaluation, there is no agreed definition of an atomic fact.
Atomic facts inherit some structure from \textit{Summarization Content Units (SCUs)} of earlier summarization evaluation work (Pyramid Method) \cite{NenkovaP04-pyramid-method}, yet remain underspecified and lack consistent formalization.
This ambiguity has practical consequences.
In our annotation experiments, we observe substantial disagreement between both human and LLM-based annotators, particularly in the amount and granularity of extracted facts.
These inconsistencies suggest a deeper methodological gap: factual correctness cannot be evaluated if the annotation ground truth disagrees with itself.
This lack of standardized definitions leads to substantial differences in fact decomposition methods, reducing the meaningfulness of LLM evaluation \cite{WannerEJDD24-claim-decomposition}.

\medskip

We argue that low inter-annotator agreement (IAA) is a signal of incomplete or ambiguous annotation guidelines.
Definitions of atomicity trade off between \textit{independence} (i.e., minimal dependency on other facts) and \textit{simplicity} (i.e., short facts that do not carry long context).
This creates structurally unstable representations: independent facts tend to become overly verbose, while simple facts often encode hidden dependencies.
Philosophical approaches like Russel's \textit{logical atomism} \cite{Russel1918-logical-atomism} or \textit{neo-Davidsonian event semantics} offer theoretical insight, yet they remain infeasible in practice due to linguistic ambiguity and severe annotation overhead \cite{GunjalD24-molecular-facts}.

\medskip

We propose to tackle this problem through visual analytics (VA):
We present a conceptual framework for a visual fact annotation tool designed to expose disagreement (see \cref{fig:teaser}), allowing for systematic annotation guideline refinement, until convergence.
Once the guidelines reach a robust state in which both humans and LLMs reach a certain level of agreement, the tool can be used as a human-in-the-loop LLM-guided annotation tool, increasing both annotation accuracy and efficiency and ultimately improving LLM evaluation.

\vfill
\clearpage

\section{Method}

\begin{figure}[t]
  \centering
  \includegraphics[width=0.49\columnwidth]{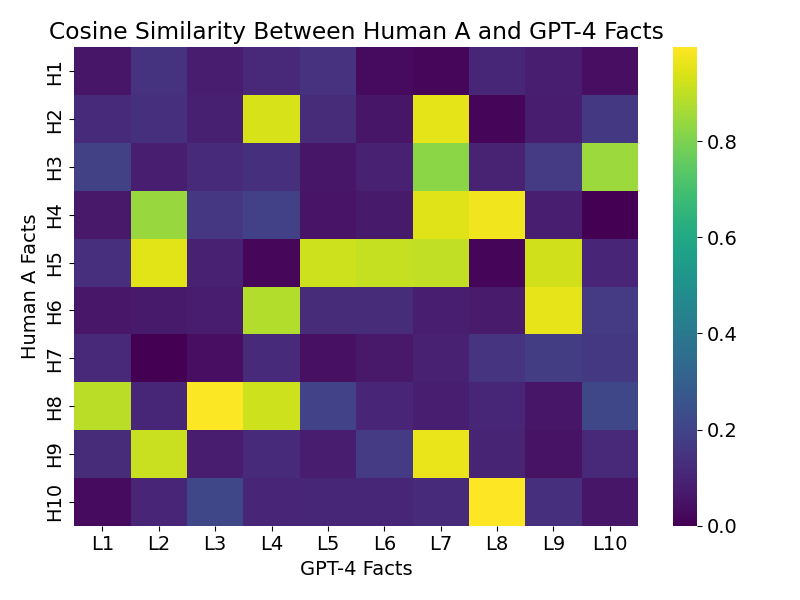}
  \hfill
  \includegraphics[width=0.49\columnwidth]{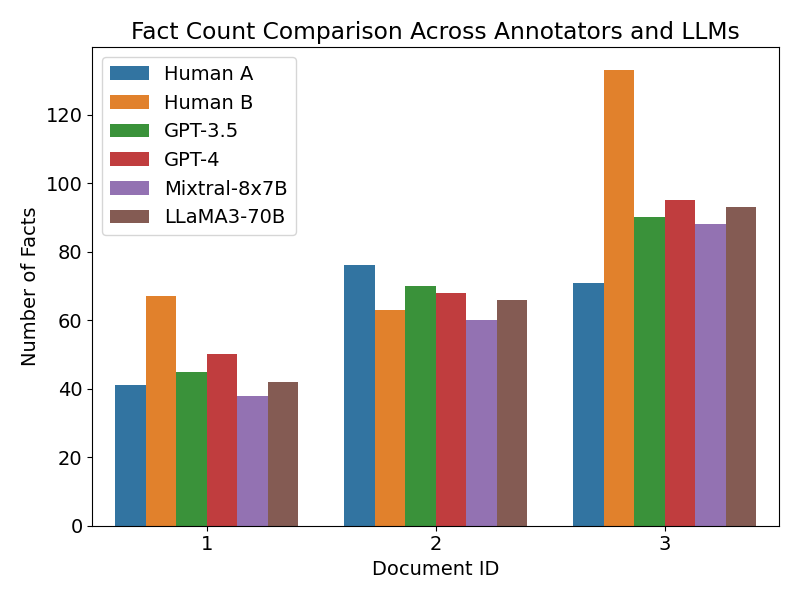}
  \caption{Initial analysis shows high disagreement between human annotators and LLMs both in semantics (left) and granularity (right).}
  \label{fig:disagreement}
  \vspace*{-1em}
\end{figure}

To study disagreements in human and LLM annotations, we manually annotated 13 German administrative documents from the platform \textit{service-bw}.
Each document was processed into a list of natural language atomic facts based on our annotation guidelines. Three of these documents were independently annotated by two human annotators to measure inter-annotator agreement (IAA).
Our annotation guidelines contain instructions on how to deal with anaphora resolution, conditionals and conjunctions.
The instruction fine-tuned LLMs (GPT-3.5, GPT-4, LlaMA3-8B/70B, Mistral-7B, Mixtral-8x7B/22B) were supplied with a plaintext version of our annotation guidelines and annotated 800 documents of \textit{service-bw}.

To compare annotations, we developed an embedding-based fact matching pipeline.
First, facts are embedded using SBERT \cite{reimers-gurevych-2019-sbert}.
Pairwise cosine similarities are computed between two fact lists.
Then, the \textit{Hungarian algorithm} is applied to find an optimal assignment of facts.
We use a similarity threshold to discard low quality matches.
The resulting IAA is the \textit{Jaccard similarity} applied to the two binary matching vectors.
The threshold was empirically optimized to minimize deviation from manual human-based matching.
We also evaluated alternative embedding methods (TF-IDF, CLIP, FlagEmbedding), but SBERT consistently aligned closest with human annotations.

Despite multiple iterations of producing detailed annotation guidelines, IAA for all human-human, human-LLM and LLM-LLM pairs remained highly variable (see \cref{fig:disagreement}, left). 
Disagreement was especially pronounced in granularity (see \cref{fig:disagreement}, right), i.e., annotators differed in how finely they decomposed conjunctive and conditional structures, and referential dependency, i.e., whether contextual elements like conditions and entities should be replicated for completeness or omitted.
Language models also diverged, often merging conceptually distinct facts, omitting conditions or overgeneralizing.
These results demonstrate the fragility of atomic fact extraction when definitions are underconstrained, indicating the need for human intervention.
To address these issues, we propose a VA system to support annotation convergence and guideline refinement.
In the following, we describe the components of our approach.

\noindent
\textbf{Text-anchored fact highlighting:} Highlights in text associated with extracted facts from multiple annotators, enabling direct comparison of extractions (see \cref{fig:teaser}, \textit{Text-View} and \textit{Facts-View}).

\noindent
\textbf{Semantic similarity heatmap:} Alignment matrix of annotator/model facts with color-coded similarities for comparative overview and detection of disagreements (see \cref{fig:disagreement}, left).

\noindent
\textbf{Fact count histogram:} Comparison of fact granularity across multiple models and annotators (see \cref{fig:disagreement}, right).

\noindent
\textbf{Knowledge graphs:} Comparison of facts and source text as small multiple entity-relation graphs to identify mismatches, uncertainties and omissions through visual aggregation (see \cref{fig:teaser}, \textit{Graph-View}).

\noindent
\textbf{Branching logic visualization:} Visualization of parsed conditionals and conjunctions into tree-structured representations to explore semantic decompositions variants.

These views help analysts identify divergent guideline interpretations, missing or ambiguous dependencies and over- or under-specified facts.
This workflow to identify disagreement is itself a four-stage loop (\cref{fig:teaser}):
(1) aggregate the plain text into a knowledge graph (KG) through entity and relation extraction to create an overview, (2) highlight these entities in the extracted facts, (3) aggregate extracted facts into small-multiple KGs and (4) project the semantics of the extracted facts back onto the original text to reveal gaps or overspecification.
Ultimately, our approach aims to support a revision loop of nested disagreement exposure, guideline refinement, reannotation, and convergence measurement.

\section{Outlook}

We are developing a prototype of the proposed VA system, implementing our conceptual design.
As the main goal is to facilitate annotation convergence, we plan to add features, such as semantic clustering of facts, and contradiction and redundancy detection.
Once annotation convergence (high IAA) is achieved, the tools will support human-in-the-loop LLM-guided annotation for greater efficiency.
LLM-based guidance will use either majority-vote based resolution of ambiguities or interactively highlight possible resolution pathways, to make the annotation process more robust and efficient.
Finally, consistent annotations will serve as a foundation for more meaningful factuality evaluation by supplying facts to methods such as FActScore.

\section{Conclusion}

Even with detailed guidelines to extract atomic facts, our empirical results show substantial disagreement among human annotators, driven by ambiguity in decomposition, i.e. dependency resolution and granularity.
We argue that such disagreement is not noise, but a signal of underspecified annotation practices.
To this end, we propose a VA-based workflow that helps to surface inconsistencies and supports guideline revision and convergence.
Our goal is enable stable and efficient factual annotation, which forms the necessary foundation for rigorous and interpretable evaluation of LLM outputs.

\acknowledgments{
We acknowledge financial support by the Federal Ministry for Economic Affairs and Climate Action (BMWK, grant 03EI1048D).
}

\bibliographystyle{abbrv-doi-narrow}

\vspace*{-0.25em}
\bibliography{bib}



\section*{Supplementary material}

# Dissecting Atomic Facts
## Visual Analytics for Improving Fact Annotations in Language Model Evaluation

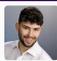 **Manuel Schmidt**
University of Konstanz
manuel.schmidt@uni.kn

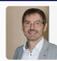 **Daniel A. Keim**
University of Konstanz
keim@uni.kn

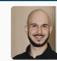 **Frederik L. Dennig**
University of Konstanz
frederik.dennig@uni.kn

## Motivation
- LLMs are prone to making up false statements
  → Hallucinations
- Tools are needed to evaluate LLM outputs for factual correctness
- Meaningful LLM Evaluation paves the way for reliable AI in sensitive domains

## LLM Evaluation using Atomic Facts
- Methods like FActScore [2] evaluate the factuality of LLMs
- Inspired by Summary Content Units from Summarization Evaluation [1]
- Utilize atomic facts
- How to match facts? Similarity based matching
- Precision $f(y) = \frac{1}{|\mathcal{A}_y|} \sum_{a \in \mathcal{A}_y} \mathbf{1}[a \text{ is supported by } \mathcal{C}]$

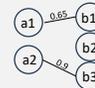

## Atomic Facts
- Smallest indivisible factual statement
  → Minimal
- Self-contained statement expressing one relation or attribute
  → Independent
- What constitues an atomic fact?

## Finding
- Difference in granularity of extracted atomic facts
  → Low inter-annotator agreement
- Annotators treat Dependency resolution, Conditionals and Conjunctions differently
- Ex.: The dog was brown **and** very tall – How many facts?

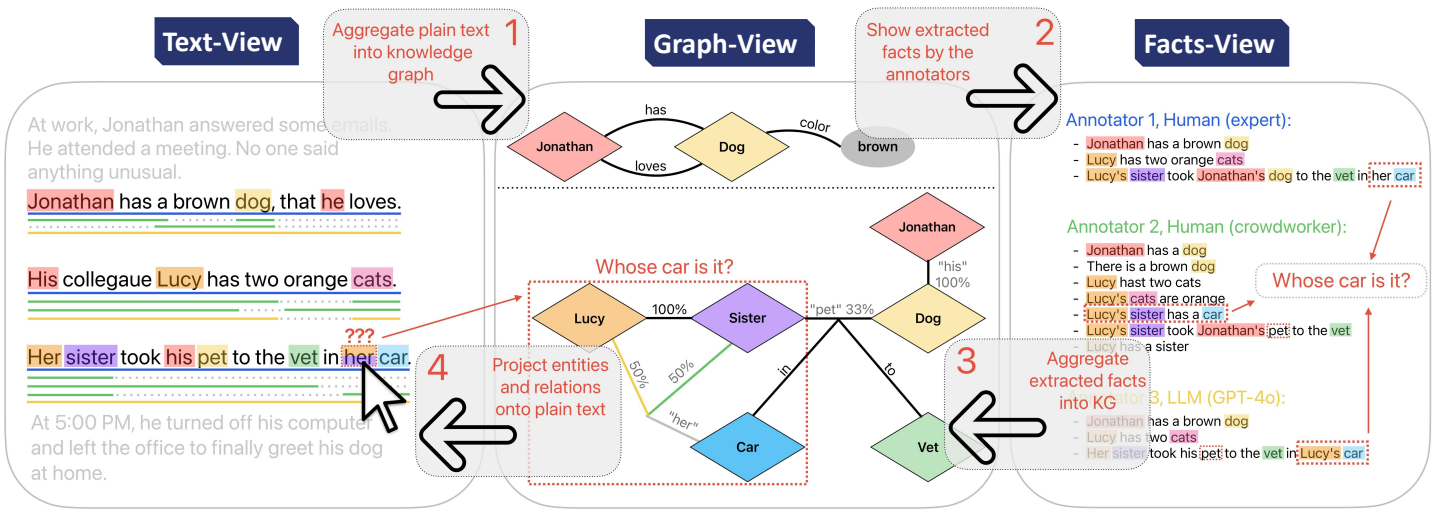

**Text-View** | **Graph-View** | **Facts-View**

1. Aggregate plain text into knowledge graph
2. Show extracted facts by the annotators
3. Aggregate extracted facts into KG
4. Project entities and relations onto plain text

- Semantic highlighting of extracted facts
- Interactive selection highlights parts in the Graph-View

- Automatic knowledge graph (KG) generation
- Create KGs from original source text and extracted facts
- Comparison through small multiples

- Shows facts extracted by human and LLM annotators
- Entity highlighting
- Ambiguity highlighting

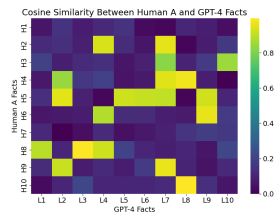
Similarity between Atomic Facts
→ Fact Matching

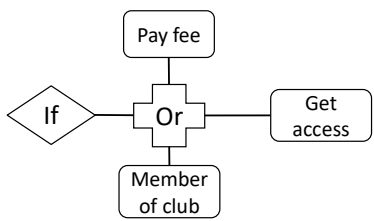
Visualize Conditionals
with Tree-Viewer

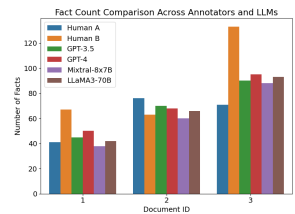
Fact Count Histogram
→ Annotator Agreement

## Outlook
- Visual Analytics Application under development
- Features such as semantic clustering help in finding ambiguities
- Contradiction and redundancy detection help to spot disagreements
- Once agreement converges: **Human-in-the-loop LLM-guided annotation**

## Conclusion
- A Visual Analytics based workflow is necessary to address the problems
- Analytic features like **detection of annotation inconsistencies** are required
- Iteratively **revise annotation guidelines**
- Goal: **Converge inter-annotator-agreement**

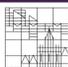
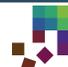



\end{document}